# The Need for Climate Data Stewardship:
# 10 Tensions and Reflections regarding Climate Data Governance

Dr. Stefaan G. Verhulst
The Governance Lab (New York)
The Data Tank (Brussels)
Center for Urban Science and Progress, New York University
ISI Foundation (Torino)
imec-SMIT, Vrije Universiteit Brussel

**Abstract:** Datafication - the increase in data generation and advancements in data analysis- offers new possibilities for governing and tackling worldwide challenges such as climate change. However, employing new data sources in policymaking carries various risks, such as exacerbating inequalities, introducing biases, and creating gaps in access. This paper articulates ten core tensions related to climate data and its implications for climate data governance, ranging from the diversity of data sources and stakeholders to issues of quality, access, and the balancing act between local needs and global imperatives. Through examining these tensions, the article advocates for a paradigm shift towards multi-stakeholder governance, data stewardship, and equitable data practices to harness the potential of climate data for public good. It underscores the critical role of data stewards in navigating these challenges, fostering a responsible data ecology, and ultimately contributing to a more sustainable and just approach to climate action and broader social issues.

___________________________

**Introduction**
We are living, as the historian Adam Tooze has argued, in the age of the polycrisis.[1] From pandemics to rampant inequality, from global warming to the rise of illiberal populism, the world faces intertwining and overlapping problems whose complexity and intractability seem to defy conventional governance methods. Increasingly, there is a sense that we need not only new solutions, but *new approaches* to solutions.

Data is often upheld as offering potential pathways: new frameworks for governance, new paradigms to help mitigate our most pressing problems. A growing chorus of experts argue that—at a time of increasing datafication, the exponential increase in data and sophisticated methods for analyzing and using it—there may be fresh avenues for governance[2] and social and economic re-ordering.[3]

---

[1] Tooze, Adam. Welcome to the world of the polycrisis. Financial Times. https://www.ft.com/content/498398e7-11b1-494b-9cd3-6d669dc3de33

At the same time, the risks of data are also becoming increasingly apparent: greater inequality and new forms of exclusion; implicit (and explicit) biases; and access asymmetries that graft themselves onto and exacerbate existing socio-economic inequities. These tensions—between the challenges and opportunities of data—are central to our age, and must be navigated by policymakers and other stakeholders seeking to address mounting crises.

In this article,[4] we examine the potential of data to address one of our most severe challenges: climate change. Climate data has often been cited as one of the most promising ways to address the vast and intertwined series of risks associated with global warming.[5] At the same time, much as with data in general, the use of climate data[6] is accompanied by certain challenges and tensions. These tensions are exacerbated by what Verhulst (2024) has called an impending "data winter"[7] —a period of decreased funding for and access to data, marked by restrictive data access policies by social media platforms, legislative and policy inaction, and the effective privatization of data. In what follows, we address ten tensions, some specifically associated with climate data and some more generally with the use of data to address social problems. Navigating these tensions, we argue, is essential to unlock the potential of climate data, and developing a framework for sustainable, systematic, and responsible use of data to address global warming and its many associated challenges. To conclude, we discuss the concept of data stewards,[8] and the key role they can play in this process, and more generally in fostering a data ecology that serves the public good.

**Tension #1: Diversity of Sources, Actors, Purposes, and Products**

---

[2] Marcucci, S., Alarcón, N. G., Verhulst, S. G., & Wüllhorst, E. (2023). Informing the Global Data Future: Benchmarking Data Governance Frameworks. Data & Policy, 5, e30.
[3] Cukier, Kenneth. The Birth of Datafication. https://bigthink.com/videos/the-birth-of-datafication/
[4] Some of the research for this article was previously used to draft
https://files.thegovlab.org/erdgovernance.pdf
[5] Leitzell, K., and N. Caud. (2021) "Climate change widespread, rapid, and intensifying–IPCC." 2021-08-09)[2021-09-27]
[6] Open Data Charter (nd) Open-Up Guide: Using Open Data to Advance Climate Action.
[7] Verhulst, S (2024) Are we entering a "Data Winter"?
[8] Data Stewards are "Individuals or teams that are empowered to proactively initiate, facilitate and coordinate data collaboratives when they may be useful or necessary. We call such individuals and teams "data stewards" Verhulst, S (2018) The Three Goals and Five Functions of Data Stewards
Data Stewards: a new Role and Responsibility for an AI and Data Age.
https://medium.com/data-stewards-network/the-three-goals-and-five-functions-of-data-stewards-60242449f378

First, the diversity of (new) climate data sources and of key actors and stakeholders involved in the climate data chain, each bringing their own priorities and values to the discussion, creates difficulties and tensions in the data ecology.

**Diversity of New Data Sources Resulting from New Methods, and Instrumentation**
- Advances in Sensing and Monitoring Technology: The Internet of Things (IoT) and new satellite technologies have significantly expanded the scope of climate data collection. IoT devices, such as sensors deployed in various environments, collect real-time data on temperature, air quality, and water levels. New satellite technologies offer high-resolution imagery and data on land use changes, deforestation, and ice melt rates.
- New Data Collection Methods: Citizen science initiatives empower individuals to contribute to data collection, using simple tools or smartphone apps to report local weather conditions, species counts, or pollution levels. This democratization of data collection diversifies and enriches the climate data ecosystem.
- Advances in Machine Learning: The development of sophisticated machine learning models, including Generative Pre-trained Transformers (GPT), has revolutionized data analysis. These models can process vast datasets to identify patterns, trends, and anomalies, making climate predictions more accurate and actionable.

**Diversity of Actors and Stakeholders**
- Institutions Executing Climate Law: International bodies (such as the United Nations), national governments, and local authorities implement and enforce climate-related regulations, relying on data to inform their policies.
- Statistical Agencies: These agencies compile and analyze environmental and climate data, producing vital statistics that inform research and policy.
- Climate Researchers: Scientists and researchers at universities and research institutes analyze climate data to understand climate change's mechanisms and impacts.
- Private Sector: Companies across various sectors use climate data for risk management, product development, and sustainability initiatives.
- Citizen Scientists: Individuals participating in data collection contribute valuable localized insights, enriching the global understanding of climate change.

**Diversity of Purposes for and Users of Climate Data**
- Science: Researchers use climate data to deepen our understanding of environmental processes and the impacts of climate change.
- Policymaking: Governments leverage climate data for informed governance, policy-making, and enforcement of environmental regulations.

> - Advocacy: NGOs and activists use climate data to advocate for climate justice and raise awareness about climate change's impacts on vulnerable populations.
> - Planning and Response: Data informs emergency response strategies, infrastructure planning, and environmental management to mitigate climate risks.
> - Economic: In sectors like agriculture, insurance, and energy, climate data supports resource management and helps develop climate-resilient business models.
>
> **Diversity of Climate Data Products**
> - Indicators: Climate indicators, such as greenhouse gas concentrations, sea level rise, and global temperature anomalies, offer concise, critical insights into the state of the climate system.
> - Statistics: Statistical analyses provide a quantitative basis for understanding trends, variations, and projections in climate data.
> - Visualizations: Maps, graphs, and interactive platforms transform complex climate data into accessible and understandable formats for diverse audiences.
> - Applications: Software and apps translate climate data into practical tools for education, decision-making, and daily life, enabling users to access personalized climate information and advice.

Box 1: Growing Diversity in the Climate Data Ecology

Box 1 indicates some of the diversity, listing a sample of data sources, actors, purposes and products involved in climate data. The diversity is only likely to increase with the continued expansion of the Internet of Things, accompanied by a plethora of sensors and other new data collection methods. In addition, climate data, once primarily collected and used by scientists, is now making its way into a variety of domains—policymaking, the private sector, disaster response, etc.—which leads to a potential divergence of methods and priorities among an ever-widening group of stakeholders (including statistical agencies, scientists, citizen scientists, policymakers, civil society, the private sector, and more). Advances in machine learning and AI are likely to further complicate the picture, leading to unpredictable uses of data and equally unpredictable outcomes.

The cornucopia of interests and stakeholders —marked both by plenty and increasing divergence — calls for new approaches to governance. In particular, there is a need for greater multi-stakeholder governance that could align interests, sensitivities, and requirements at all levels of decision making.[9] Such governance would help ensure legitimacy and trust, essential ingredients in a field that often processes PII and where higher standards of care and responsibility are required. Multi-stakeholder governance can also help protect independence and rigor for scientists and other researchers, a vital concern in the field of climate data.

---
[9] Verhulst, S. (2016). The Practice and Craft of Multi-Stakeholder Governance. The Practice and Craft of Multi-Stakeholder Governance, Global Partners Digital.

**Tension #2 Competing concerns and lack of common principles**
Among the divergence of sources and interests marking the climate data field, few are as pronounced as those separating climate change and climate justice actors on the one hand and private sector stakeholders on the other. The former category (which includes policymakers, researchers, and activists) calls for more data related to the environment, more openness (embodied by the "Right to Know" movement), and seeks to increase data justice while limiting data extraction and data colonialism. Corporations and private actors, on the other hand, are often motivated by a desire to maintain competitive advantage. In addition, the private sector is itself marked by competing interests, depending on whether companies primarily produce, process, or reuse data.

In short, the field is marked by an absence of common principles for how data should be (re)used, with different stakeholders upholding different principles to advance their respective agendas and priorities. There is an urgent need for a common normative and ethical framework that could guide the collection, processing and (re)use of climate data. In particular, we need to move from a concept of data ownership, which exacerbates asymmetries[10], to data stewardship. Data stewards could increase accessibility and transparency while accommodating individual and collective concerns and rights, making room for the variety of stakeholders.

**Tension #3 Power imbalance: Who decides what to measure? And what to collect? And what to share?**
Decisions concerning what data to collect, analyze, use—and how to use it—underlie another tension at the heart of climate data. The divergence of stakeholders means similar divergence in approaches to measuring, collecting, and sharing data. The potentially deleterious impact of these divergences is heightened by power asymmetries within the digital ecology, particularly as they relate to data access and agenda-setting. An agenda for data-driven collaboration can "inform the strategic allocation of resources for new research projects, indicators for regular monitoring, and the formation of cross-sectoral data-sharing collaborations[11]".Despite clear potential benefits and widespread agreement on the general principle of using climate data, in other words, there remain important open questions about what specific forms of data are used, what types of conclusions are derived, and who benefits from data-driven decision-making.

---

[10] Verhulst, S. G. (2024). The ethical imperative to identify and address data and intelligence asymmetries. AI & SOCIETY, 39(1), 411-414.
[11] Verhulst, S., Bustamante, C. M. V., Carvajal-Velez, L., Cece, F., Requejo, J. H., Shaw, A., ... & Zahuranec, A. J. (2023). Toward a demand-driven, collaborative data agenda for adolescent mental health. Journal of Adolescent Health, 72(1), S20-S26.

Solutions do exist, but it is important to understand that the choice—and application—of solutions is itself not value neutral. The notion of purpose specification, for instance, which may help limit data abuses, is an inherently political choice: defining a purpose will inherently narrow the scope of who benefits. To help mitigate such risks, it is essential that equity be considered as an overarching principle throughout the data value chain and the broader ecology of climate data governance.

**Tension #4 Extraction through Collection. Proportionality and Collective rights. Data ownership**

Datafication is taking place in a world marked by long-standing hierarchies, inequalities, and socio-economic divisions. This is perhaps especially true of climate data, which is being collected, processed and used across national and cultural boundaries, raising important questions about relations between and the relative rights of communities. Concerns exist about data sovereignty[12] for indigenous populations, feminist and anti-colonial movements, and the rights of populations subject to so-called "helicopter research"[13] and extractive data practices. Such concerns are magnified by a general distrust of data collection practices between populations with unequal rights or power[14], and growing awareness regarding flawed data consent provisions, which call for new "social licenses" to govern how data is collected and used.

To ensure that the field of climate data mitigates rather than exacerbates existing divisions, such concerns should be acknowledged by and embedded within emerging governance frameworks. Community-based participatory research and collection can help minimize extractive data practices. Representation should also be taken into consideration when selecting data stewards, whether they be individuals or bodies and institutions (which can be composed of a variety of community voices so as to ensure various stakeholder interests are taken into account).

In addition, it is essential to uphold proportionality as a core principle, for example by tailoring data collection efforts to the minimal necessary for meaningful insights, ensuring that the benefits of data use are equitably distributed among all stakeholders, and implementing safeguards that prevent the over-surveillance of vulnerable communities. Underlying these methods and approaches is a recognition of data as a public—rather than private—good, and a commitment to upholding the principle of digital self-determination[15] throughout the data ecology.

---

[12] Verhulst, S. G. (2023). Operationalizing digital self-determination. Data & Policy, 5, e14.
[13] Tackling helicopter research. Nat. Geosci. 15, 597 (2022). https://doi.org/10.1038/s41561-022-01010-4
[14] Verhulst, S., & Young, A. (2022). Identifying and addressing data asymmetries so as to enable (better) science. Frontiers in Big Data, 5, 888384.

**Tension # 5 Quality, Provenance, and Standards**
As the volume of available data grows, so do concerns over data quality and integrity. In addition, it is essential to ensure—and acknowledge—the "situatedness" of data; so-called "thick data"[16] helps ensure that data is processed, analyzed and used in a contextually relevant and sensitive way. Attention to data quality and thickness must be accompanied by—and can help encourage—the development of data standards to promote interoperability and responsible reusability. Responsible reuse is especially important for climate data, which is inherently global and cross-national in scope.

Data quality is in part a matter of ensuring that decisions involving climate technology are supported by appropriate quality-assured engineering standards and processes. But, as with all the tensions discussed in this paper, resolving these tensions is about far more than just technology. Governance and policy steps are also essential. These can include frameworks that help operationalize and standardize quality assurance, for instance, by establishing clear metrics for data accuracy, consistency, and completeness across different stages of data collection, analysis, and usage. In addition, data quality (and relevance) can be enhanced if institutions have policies that track what decisions they make about data throughout the data lifecycle. Finally, treating carbon as a currency puts pressure throughout the system to prioritize transparency, accountability, and accuracy in data reporting, thereby ensuring that climate action efforts are both effective and verifiable.

**Tension #6 Timeliness, Continuity and Sustainability**
Ensuring timely collection and release of climate data is critical for effective risk management and the development of innovative approaches to advance climate mitigation and climate justice. However, persistent concerns regarding the financial (and other forms of) sustainability raise questions about timeliness, continuity and the long-term viability of climate data and efforts at climate justice. Such concerns are heightened by the political context of the global climate debate, which poses challenges to policy continuity and financial sustainability. In particular, a risk exists that bigger stakeholders could walk away from the conversation, thus imperiling the functioning of the larger ecosystem dedicated to climate mitigation.

---

[15] International Network on Digital Self-Determination. https://www.idsd.network/
[16] Thick Data vs. Big Data. https://www.bbvaopenmind.com/en/technology/digital-world/thick-data-vs-big-data/

As a result of these tensions, it is clear that any efforts at sustainability need to be at the core of an effective governance framework for climate data. This will require fit-for-purpose incentives for investment and institutionalization so that all stakeholders are aligned. In addition, governance will need to ensure the timeliness of climate data, to ensure its relevance and effectiveness. Some mechanisms that can help achieve these objectives include public-private data collaboratives[17][18] (e.g., the Net Zero Public Data Utility[19]), standards for data quality and timeliness, and international agreements on climate data exchange and reporting. In addition, fostering community-based monitoring programs and leveraging technological innovations, such as artificial intelligence for data analysis, can further enhance the governance framework's effectiveness.

**Tension # 7 Access, Openness, and Transparency**
Our era's challenges necessitate nuanced solutions beyond merely "opening" data—a narrow approach that may inadvertently serve corporate agendas. Mere access to raw data does not guarantee transparency, leaving room for manipulation. To navigate these complexities, fostering data collaboratives, empowering data stewards, and enhancing data sharing practices are essential. However, for these strategies to be genuinely effective, it is crucial to establish and clarify incentives for private sector data holders ("the business case" for data collaboration[20]) and to improve the harmonization of cross-border data sharing efforts.

As a broad approach, embedding FAIR data principles[21] into the climate data conversation may provide an overarching framework to reduce access asymmetries and achieve meaningful openness. These principles (which embody findability, accessibility, interoperability, and reusability) were first established in the context of academic data. In addition, data collaboratives have proven particularly useful vehicles for bridging gaps between the private and public sectors, and more generally for promoting greater data sharing. And finally, the field of climate data could benefit from a commitment by relevant stakeholders to share data in response to specific—and specific types or scales—of crises. This can at least ensure that the right data will be available to serve the public good when it is most needed.

---

[17] Data Collaboratives. https://datacollaboratives.org/
[18] Verhulst, S. G. (2023). Data collaboratives and data sharing. Internet Sectoral Review (4:15).
[19] Net-Zero Data Public Utility https://nzdpu.com/home
[20] Zahuranec, A, Young, A & Verhulst. S (2021) The "9Rs Framework": Establishing the Business Case for Data Collaboration and Re-Using Data in the Public Interest/
[21] Wilkinson, Mark D., Michel Dumontier, IJsbrand Jan Aalbersberg, Gabrielle Appleton, Myles Axton, Arie Baak, Niklas Blomberg et al. "The FAIR Guiding Principles for scientific data management and stewardship." *Scientific data* 3, no. 1 (2016): 1-9.

**Tension # 8 Bias, Capture and Whitewashing**
Tension #3, above, highlights the risks of power imbalances within the data ecology. There are various ways these imbalances can manifest. In addition to the problems posed by unequal access, bias in the underlying data and algorithms is emerging as a serious concern—particularly given the growing prominence of artificial intelligence and large language models (LLMs). These risks are accompanied by concerns over "whitewashing" (in which stakeholders may cover up or obfuscate data collection or disclosure), and concerns regarding academic capture[22] by the private (and public) sectors. All of these affect the rigor and credibility of data and data efforts; more generally, they undermine trust in the ecosystem.

An adequate framework to address these concerns must encompass both technological solutions and policy interventions. Technological solutions might involve the development of automated systems capable of auditing and tracing data flows within the lifecycle, thereby ensuring accountability through alerts on potential discrepancies or misuse. However, these technological measures need to be underpinned by a robust governance framework that not only operationalizes ethical principles but also extends the notion of responsibility in the use of climate data.

This broader governance framework should include comprehensive approaches to protect individual and community rights. Beyond these, it is essential to establish a "social license to operate[23]," which entails gaining and maintaining public trust by demonstrating that data collection and use are conducted in ways that align with societal values and expectations. This concept goes beyond legal compliance to include ethical considerations, transparency, and community engagement, ensuring that data practices are perceived as legitimate and beneficial by the wider community.

**Tension # 9 Local vs Global: Subsidiarity and Cultural difference**
Climate change is a global problem, and climate data is therefore also global in scope. But within this broad context, there exists little consensus on what types of

---

[22] Academic Capture. Regulatory Capture Lab https://regulatorycapturelab.ca/Academic-Capture
[23] Verhulst, S, Sandor L & Stamm J (2023) The Urgent Need to Reimagine Data Consent.
https://ssir.org/articles/entry/the_urgent_need_to_reimagine_data_consent

questions[24][25]—or solutions—should be devolved to the local level, and how to account for cultural and structural differences across jurisdictions. As a general principle, governance at the local level is more conducive to citizen participation, and rapid feedback loops to iterate and improve upon data initiatives. In addition, local governance allows for more efficient management of climate resources and challenges. At the same time, it is important to recognize that many localities may lack the type of data expertise available at national and international levels. An overly local focus may likewise lack the global perspective required to mitigate broader challenges.

Several pathways are available to resolve these tensions. To begin, it is essential to embed notions of subsidiarity within data governance design; these can help establish the principle that, where possible and advantageous, climate data should be governed at the local level. Highlighting subsidiarity also upholds the importance of local participation, and cultural and context sensitivity. Alongside subsidiarity, however, it is equally important to identify issues and challenges that are better addressed at the global (or supra-local) level—for example, data related to the management of common resources or the upholding of shared values and rights. Ultimately, an effective governance framework will coordinate the local and the global, and harmonize the need for contextual sensitivity and community participation with the wider scope of the problem and of the data itself.

**Tension #10 Disputes, Accountability and Use**
While it is increasingly clear that data can help mitigate the climate crisis, there exists little agreement on how to implement it, and what processes or mechanisms exist to harmonize values and priorities, avoid misuse and harm, and ensure accountability. Disputes are inevitable, and require agile and independent processes to be resolved in a productive manner. In addition, as with all uses of data, accountability is essential. Climate justice requires not just a judicious use of data but also clear lines of responsibility and accountability.

Existing processes from other data verticals may provide guidance. For example, there now exist well-established procedures and mechanisms (technological and otherwise) to establish decision provenance (i.e., transparency about who is responsible and

---

[24] The 100 Questions Initiative. The Govlab. https://thegovlab.org/project/project-the-100-questions-initiative

[25] Verhulst, Stefaan. "Questions as a Device for Data Responsibility: Toward a New Science of Questions to Steer and Complement the Use of Data Science for the Public Good in a Polycentric Way." *Aguerre, C., Campbell-Verduyn, M., & Scholte, JA, Global Digital Data Governance: Polycentric Perspectives, Properties and Controversies. Routledge, Forthcoming* (2023).

accountable for the use of climate data). Auditing tools and frameworks can also be repurposed, designed to ensure independence and agility within dispute resolution processes. While repurposing such established steps, it is also important to keep in mind specific needs and variations that may be necessary in the climate context—e.g., the tensions between the global and the local or the urgent need for real-time data to inform immediate climate action, the necessity for integrating indigenous and traditional knowledge systems, and the importance of addressing data sensitivity and security concerns related to vulnerable ecosystems and communities

**Conclusion: Toward A Decade of Data and Climate Data Stewardship**

The preceding has outlined ten tensions that currently characterize the climate data ecology. Our ability to use data for the public good—and more generally to mitigate the impending climate crisis—depends significantly on the extent to which we are able to navigate these tensions productively, maximizing the benefits of climate data while limiting their potential harms. At the same time, even as we seek to navigate the specific tensions, we must also ensure a broader commitment to data access, and nurture a data ecology that fosters responsible reuse of private data toward the public good.

What's required, in effect, is an International Decade of Data[26]. This decade would be marked by new activity on the legislative front as well as a general cultural shift in how society views data and the ability to reuse private data for the public good. This requires awareness raising and capacity building, and it requires stakeholders from various sectors and from around the world—this is both a global and local problem—to come together to limit data hoarding and instead foster responsible sharing and reuse.

Key in that endeavor must be the advancement of climate data stewardship. Data stewards can play a key role in this task, and in helping to foster a more responsible climate data ecology. In particular, data stewards play a key role in fostering data collaboratives and greater data access and reuse.

Within the field of climate data, data stewards have three key responsibilities:

- **Collaborate:** Data stewards have a responsibility to identify, nurture and manage data and data collaboratives when there is an opportunity to unlock data in the public interest. As part of this responsibility, data stewards can help break down data stored in private silos.

---

[26] Verhulst, S. (2023). Unlocking the Potential: The Call for an International Decade of Data. https://unu.edu/publication/unlocking-potential-call-international-decade-data

- **Protect:** Data stewards play a key role in managing data ethically and preventing harm and misuse. In this role, data stewards also help protect the integrity and quality of data.

- **Act:** Finally, data stewards have a responsibility to proactively help unlock data and–this is critical–ensure that data insights are acted upon responsibly and in the public interest.

Table 1 breaks down these broad categories, showing some specific steps that data stewards can take to unlock the ten tensions discussed in this paper. Considered together, the items in the table offer an action list for the climate field, akin to a SOW (scope of work) for prospective data stewards working for climate justice and to mitigate climate change through more responsible use of data.

| 10 Tensions in Climate Data Governance | 10 Steps toward Climate Data Stewardship |
|---|---|
| 1. Diversity of Sources, Actors, Purposes, and Products: The vast array of data sources, stakeholders, and applications leads to complexity in data governance.<br>2. Competing Concerns and Lack of Common Principles: Different stakeholders uphold varying principles for data use, necessitating a unified framework for climate data governance.<br>3. Power Imbalance: Decisions about data collection and use are influenced by disparities in power among stakeholders.<br>4. Extraction through Collection: Issues of data sovereignty, particularly for marginalized communities, raise concerns about equitable data practices.<br>5. Quality, Provenance, and Standards: Ensuring data integrity and developing | 1. Foster Multi-Stakeholder Governance: Encourage collaboration among diverse stakeholders to align interests and standards.<br>2. Develop a Normative Framework: Create a common ethical framework guiding the collection, processing, and use of climate data.<br>3. Enhance Equity: Incorporate equity considerations across the data value chain to address power imbalances.<br>4. Promote Data as a Public Good: Recognize data as a resource for the collective benefit, ensuring access and equitable distribution of insights.<br>5. Implement Data Quality and Integrity Measures: Prioritize the accuracy, reliability, and contextuality of data |
| standards for interoperability are essential for reliable climate action. 6. Timeliness, Continuity, and Sustainability: The need for timely data collection conflicts with challenges in sustainability and policy continuity. 7. Access, Openness, and Transparency: Balancing the need for open data with the protection of sensitive information is crucial for trust and effectiveness. 8. Bias, Capture, and Whitewashing: Addressing biases in data and algorithms is vital for credible and inclusive climate action. 9. Local vs. Global: Navigating the tension between local autonomy and global coordination is key for effective climate governance. 10. Disputes, Accountability, and Use: Establishing mechanisms for resolving disputes and ensuring accountability is critical for the ethical use of climate data. | through robust quality assurance frameworks. 6. Ensure Sustainability: Develop incentives and policies that support the long-term viability of climate data initiatives. 7. Embrace Data Collaboratives and Adopt FAIR Data Principles: Apply principles of Findability, Accessibility, Interoperability, and Reusability to improve data access and use. 8. Mitigate Biases and Foster Transparency: Utilize technological and policy measures to address data biases and promote openness. 9. Balance Local Autonomy with Global Coordination: Leverage the principle of subsidiarity to empower local decision-making while ensuring global coherence. 10. Establish Dispute Resolution and Accountability Mechanisms: Create independent processes for addressing disputes and ensuring responsible data use. |

**Table 1: Tensions and Steps toward Climate Data Stewardship**

Today, the world stands at the precipice of a major crisis that has huge — even existential — implications. Data is of course not a silver bullet; technology cannot solve the climate crisis on its own. But data is an essential component of any solution. We owe it to future generations (and to ourselves) to seize the moment by recommitting ourselves to the ethical use and reuse of data, and to creating a more just and equitable data ecology. The road to climate justice runs, at least in part, through data justice.